\documentclass[journal]{IEEEtran}
\usepackage{stfloats}

\usepackage{enumerate}
\usepackage{algorithm,algpseudocode}
\usepackage{setspace,amsmath,latexsym,cite,amssymb,epsfig,amsfonts}
\usepackage{url,cite}
\usepackage{graphicx}
\usepackage{psfrag}
\usepackage{footmisc}
\usepackage{multirow}
\usepackage{color}
\usepackage{multicol}
\usepackage{mathtools}
\usepackage{subcaption}
\usepackage{graphicx}
\usepackage{amssymb}
\usepackage{amsmath}
\usepackage{epstopdf}
\usepackage{geometry}
\usepackage{float}
\usepackage{balance}
\usepackage{makecell}
\usepackage{changepage}

\algnewcommand\algorithmicforeach{\textbf{for}}
\algdef{S}[FOR]{For}[1]{\algorithmicforeach\ #1\ \algorithmicdo}

\twocolumn

\makeatother

        \makeatletter
        \def\fps@eqnfloat{!t}
        \def\ftype@eqnfloat{4}
        
        \newenvironment{eqnfloat*}
               {\@dblfloat{eqnfloat}}
               {\end@dblfloat}
        \makeatother
\allowdisplaybreaks[4]

\title{Flexible Reconfigurable Intelligent Surface-Aided Covert Communications in UAV Networks}
\author{Chong Huang, \IEEEmembership{Member, IEEE}, Gaojie Chen, \IEEEmembership{Senior Member, IEEE}, Zhuoao Xu, \IEEEmembership{Member, IEEE},\\Jing Zhu, \IEEEmembership{Member, IEEE}, Taisong Pan, Rahim Tafazolli, \IEEEmembership{Fellow, IEEE}, and Wei Huang
\thanks{C. Huang, Z. Xu and R. Tafazolli are with 5GIC \& 6GIC, Institute for Communication Systems (ICS), University of Surrey, Guildford, GU2 7XH, United Kingdom. Email: chong.huang@surrey.ac.uk, zhuoao.xu@ieee.org, r.tafazolli@surrey.ac.uk}
\thanks{G. Chen is with the School of Flexible Electronics (SoFE) \& State Key Laboratory of Optoelectronic Materials and Technologies, Sun Yat-sen University, Guangdong, China, and also with 5GIC \& 6GIC, Institute for Communication Systems (ICS), Home for 5GIC \& 6GIC, University of Surrey, Guildford, GU2 7XH, United Kingdom. Email: gaojie.chen@ieee.org (Corresponding author: G. Chen)}
\thanks{J. Zhu and W. Huang are with the School of Flexible Electronics (SoFE) \& State Key Laboratory of Optoelectronic Materials and Technologies, Sun Yat-sen University, Guangdong, 510220, China. Email: zhuj229@mail.sysu.edu.cn, huangw323@mail.sysu.edu.cn.}
\thanks{T. Pan is with the School of Materials and Energy, University of Electronic Science and Technology of China, Chengdu 610054, China. Email: tspan@uestc.edu.cn.}
}

\begin{document}
\captionsetup[figure]{name={Fig.},labelsep=period}

\begin{singlespace}
\maketitle

\end{singlespace}

\thispagestyle{empty}
\begin{abstract}
In recent years, unmanned aerial vehicles (UAVs) have become a key role in wireless communication networks due to their flexibility and dynamic adaptability. However, the openness of UAV-based communications leads to security and privacy concerns in wireless transmissions. This paper investigates a framework of UAV covert communications which introduces flexible reconfigurable intelligent surfaces (F-RIS) in UAV networks. Unlike traditional RIS, F-RIS provides advanced deployment flexibility by conforming to curved surfaces and dynamically reconfiguring its electromagnetic properties to enhance the covert communication performance. We establish an electromagnetic model for F-RIS and further develop a fitted model that describes the relationship between F-RIS reflection amplitude, reflection phase, and incident angle. To maximize the covert transmission rate among UAVs while meeting the covert constraint and public transmission constraint, we introduce a strategy of jointly optimizing UAV trajectories, F-RIS reflection vectors, F-RIS incident angles, and non-orthogonal multiple access (NOMA) power allocation. Considering this is a complicated non-convex optimization problem, we propose a deep reinforcement learning (DRL) algorithm-based optimization solution. Simulation results demonstrate that our proposed framework and optimization method significantly outperform traditional benchmarks, and highlight the advantages of F-RIS in enhancing covert communication performance within UAV networks.
\end{abstract}


\section{Introduction}
In this section, we introduce the background of flexible reconfigurable intelligent surface (F-RIS), UAV networks and covert communications, along with related works. Then, we will summarize the motivations and contributions of this work.

\subsection{Background}
With the development of wireless communications, in scenarios such as the Internet of Things (IoT) and unmanned aerial vehicle (UAV) networks, the high level requirements of privacy and security demands on security wireless transmission technologies \cite{9946859}. Traditional encryption methods, e.g., symmetric or asymmetric encryption, primarily focus on protecting the confidentiality of information but do not conceal the existence of communication itself. Even if the transmitted data is protected, the communication activity can still be detected. Physical layer security (PLS) improves transmission security by using the fundamental ability of the physics of radio propagation, but communication activities are still detectable. In contrast, covert communication aims explicitly at hiding the existence of communication by embedding signals into background or artificial noise, making it difficult for third-party observers to detect ongoing transmissions, thereby offering a higher level of privacy protection \cite{10090449}.

In future wireless communication networks, UAV communications have emerged as a promising technology due to its flexible deployment and strong adaptability \cite{7470933}. However, UAVs are frequently required to transmit or receive sensitive information, and face significant risks from malicious attacks. Even with advanced encryption methods and PLS algorithms, malicious attackers may still obtain confidential information through methods such as password cracking or analysis of signals and channels. Unlike encryption and PLS approaches which are designed to prevent data eavesdropping, covert communications focus on concealing the existence of the communication itself to meet stricter security requirements. Moreover, compared to terrestrial communications, the mobility of UAVs can be better integrated with covert communication techniques to significantly reduce the probability of detection by third-party observers \cite{9382022}.

Reconfigurable intelligent surfaces (RIS) have emerged as a promising technology in wireless communications \cite{9475160}. RIS can dynamically adjust the characteristics of electromagnetic waves by introducing reconfigurable reflection elements into wireless transmissions, thereby reducing interference and enhancing transmission efficiency \cite{10499205}. In UAV covert communications, RIS can adjust its elements reflection phases to enhance the signal strength for legitimate receivers without being detected by eavesdroppers, this helps enhance both security and transmission efficiency of wireless communications \cite{9943536}. However, traditional RIS employs rigid materials \cite{metafilm,Wavefront}, this design limits its deployment flexibility and adaptability in complex environments, especially in emergency communication scenarios such as disaster recovery or remote field operations. In recent years, F-RIS have emerged as an innovative technology capable of flexibly manipulating electromagnetic waves through reflective elements with adjustable phases integrated onto lightweight, bendable substrates \cite{li2025flexible}. Compared to traditional rigid RIS, F-RIS offers advanced deployment flexibility, it can adapt to complex and dynamic environments by conforming to curved surfaces or irregular structures, thus significantly extending the coverage range and enhancing the performance of wireless communications. By intelligently adjusting reflection coefficients and its own physical shape, F-RIS can effectively reconfigure wireless propagation environments to enhance the security performance of wireless transmissions. Especially in highly dynamic scenarios such as UAV communication networks, F-RIS can provide high deployment adaptability and precise control over signal propagation, this demonstrates a promising prospect for applications in future wireless communications.

\subsection{Related Work}\label{sec:RW}
Secure transmissions in wireless networks have consistently been a challenge in future communication research works. Due to its distinctive security mechanism, covert communications have emerged as a crucial research direction for secure wireless networks \cite{7355562,10292916,10572320,10854524}. In covert communications, the transmitter conceals transmitted signals within the noise to prevent the detection of signal transmission activity by third parties. Numerous related works have already been studied in this area \cite{9612181}. In \cite{9612181}, an overt channel selection rule was studied to maximize the eavesdropper's detection error probability (DEP). To maximize the covert communication rate, a joint cooperative jamming and relay selection strategy was proposed for wireless cooperative networks in \cite{10292916}. In \cite{10572320}, DEP was studied under perfect channel state information (CSI) and statistical CSI scenarios in cognitive networks, and the corresponding covert rate was maximized by using iterative search algorithms. To maximize the received signal-to-noise ratio (SNR) for covert communications, an alternating optimization algorithm was studied to optimize the transmit power and pre-coding vectors for multi-antenna cooperative networks in \cite{10854524}.

The openness of UAV communications makes them easy to face eavesdropping, interference, and spoofing attacks from malicious users, this leads to serious risks for secure transmissions within UAV networks \cite{8675384}. Therefore, covert communication techniques for UAV networks have become an effective solution to enhance the security performance and privacy protection. In \cite{9830039}, the transmit power and block-length were optimized to enhance the throughput in UAV relaying covert communications. Convex optimization algorithms were studied to maximize the average covert rate by optimizing the transmit power and UAV trajectory in UAV covert communications \cite{10418490}. In \cite{10938035}, closed-form expressions of covert communication metrics were investigated for multi-antenna users in UAV communications, and the average covert rate was maximized by using convex optimization methods. To maximize the worst-case covert rate in ambient backscatter communications \cite{10949220}, a convex optimization method was utilized to adjust the transmit power and UAV deployment in UAV relay networks. In \cite{11122485}, the average covert probability and connection outage probability was studied to analyze the covert communication performance in UAV communications. The UAV's trajectory and transmit power were jointly optimized to enhance the covert communication performance in \cite{11104958}.

RIS can dynamically configure the electromagnetic environment to meet different signal transmission requirements, it has emerged as a key technological approach in covert communications \cite{9108996}. In \cite{10375580}, RIS was utilized to enhance the covert rate by resource allocation optimization strategies, and the corresponding closed-form expressions were derived. To study the performance of simultaneous transmitting and reflecting RIS (STAR-RIS) in covert communications, a convex optimization based strategy was utilized to maximize the covert rate in \cite{10478948}. In \cite{10923651}, the outage probability, intercept probability, and DEP were derived in RIS-assisted covert communications, and a convex optimization-based power allocation strategy was studied to maximize the security performance. An alternative optimization method was investigated to optimize the covert communication performance in active STAR-RIS assisted integrated sensing and communication (ISAC) systems \cite{10974475}. Moreover, RIS can dynamically adjust reflection vectors to optimize signal paths in UAV communications, this effectively cooperates with the UAV mobility to achieve joint optimization in three-dimensional (3D) space \cite{10287142}. In \cite{9943536}, an alternative optimization of transmit power, RIS phase shifts and the UAV deployment was proposed to maximize the covert rate in UAV-RIS networks. To investigate the average covert rate of worst-case in covert communications, a joint beamforming, RIS phase shift and UAV trajectory optimization was studied in \cite{10546987} under uncertain CSI and perfect CSI scenarios. In \cite{10680088},  an alternating optimization of beamforming, RIS reflection vectors and UAV deployment was proposed to maximize the covert rate in STAR-RIS-aided UAV networks. However, in typical UAV-based emergency communication scenarios, environment conditions usually cannot allow the deployment of traditional rigid-material RISs. Thus, RIS which can be used for flexible deployment and adaptation to various shape requirements are needed in UAV covert communications.

\subsection{Motivation and Contributions}
This paper investigates F-RIS-aided covert communications among UAVs. Due to the ability of conforming to structures with varying geometries, F-RIS can be deployed flexibly on buildings or vehicles to meet the challenging environmental demands of emergency communications. Furthermore, compared to traditional RIS, F-RIS can further reconfigure the electromagnetic environment by dynamically altering its shape. Moreover, we introduce non-orthogonal multiple access (NOMA) technology among UAV communications and joint UAV trajectory optimization to enhance the covert transmission rate while ensuring covert constraints are satisfied. Considering the complexity of jointly optimizing UAV trajectories, F-RIS reflection vectors and deformations, as well as power allocation, we utilize a state-of-the-art deep reinforcement learning (DRL) algorithm to address this complicated non-convex optimization problem. The main contributions of this paper are summarized as follows:

\begin{itemize}
  \item We propose a covert communication framework utilizing F-RIS to enhance covert transmission rate among UAVs. F-RIS's flexible and adaptive shape deformation capability is used to conform to curved surfaces or irregular structures, this ability can further dynamically reconfigure the electromagnetic environment to solve the demands of emergency UAV communications scenarios. This research establishes an electromagnetic model for F-RIS, where the interaction of a parallel beam with the surface is treated as a combination of beams incident at varying angles on individual RIS units. It further develops a fitted model that describes the relationship between reflection amplitude, phase, and incident angle, offering valuable insights into the angle-dependent reflection properties of F-RIS.

  \item To further improve covert transmission rate with covert constraints, we integrate non-orthogonal multiple access (NOMA) technology into UAV communication systems. This integration allows UAVs to share spectral resources efficiently while maintaining covert communication requirements.

  \item The proposed problem includes the joint optimization of multiple variables: UAV trajectories, F-RIS reflection vectors and shape deformations, and power allocation. Considering this is a complicated non-convex optimization problem, we employ an advanced DRL algorithm to adaptively address these intertwined optimization variables.

  \item We analyze the impact of using F-RIS in UAV covert communications through simulations. The results show improvements in covert transmission rates through jointly optimizing resource allocation in F-RIS-aided UAV networks, and also present the impact trends of F-RIS deformation on covert transmission rate. This study provides theoretical support for the deployment of RIS-assisted covert communications in emergency scenarios and remote areas.
\end{itemize}

The rest of this paper is summarized as follows: Section \ref{sec:sm} introduce the system model of the F-RIS-aided UAV covert network. The analysis of electromagnetic wave interactions in F-RIS and problem formulation are provided in \ref{sec:FRIS}. In Section \ref{sec:DRL}, a DRL-based optimization method is presented. Section \ref{sec:sim} provides the analysis of the proposed system and comparisons with different conditions. Finally, Section \ref{sec:con} concludes this paper.

\section{System Model } \label{sec:sm}
\begin{figure}[t!]
  \centering
  \centerline{\includegraphics[width=0.4\textwidth]{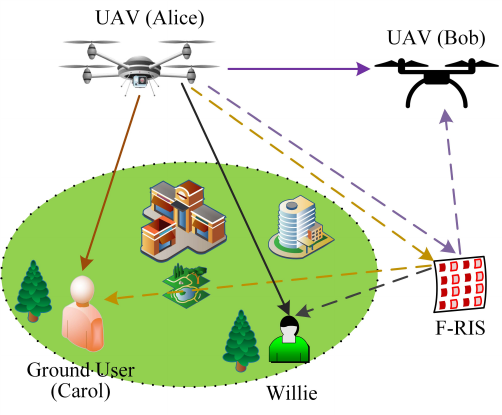}}
 \caption{Covert communications in F-RIS-aided UAV networks..} \label{fig:SM}
\end{figure}

As shown in Fig. \ref{fig:SM}, we consider an F-RIS-aided UAV network for covert communications, where a UAV (Alice) simultaneously performs covert communications with another UAV (Bob) and public transmissions to a ground user (Carol), a Willie keeps monitoring the wireless transmissions from Alice. The 3D coordinates of Alice and Bob at time slot $t$ are $q_a(t) = \{q^a_x(t), q^a_y(t), q^a_z(t)\}$ and $q_b(t) = \{q^b_x(t), q^b_y(t), q^b_z(t)\}$, respectively. To enable covert communications, Alice employs the public signal intended for Carol as a cover for the transmission to Bob. An F-RIS with $M$ reflection elements is deployed in this scenario to enhance the transmission performance. The channel coefficients between UAVs are assumed to experience line-of-sight (LoS) fading \cite{9209992}, $h_{a,b} = \sqrt{\gamma_0} d^{-{\alpha_L}/2}$, where $\gamma_0$ denotes the channel gain at the reference distance 1 m, $\alpha_L$ is the path loss exponent for LoS channels. The channels between UAV and ground user Carol follow Rician fading as \cite{10440193}

\begin{equation}\label{rician}
h_{a,c}= \sqrt{\frac{\kappa}{\kappa+1}} \bar{H}_{a,c}+\sqrt{\frac{1}{\kappa+1}} \hat{H}_{a,c},
\end{equation}
where $\kappa$ is the Rician factor, $\bar{H}_{a,c} = \bar{\mathbb{g}}_{a,c} d_{a,c}^{-{\alpha_L}/2}$ and $\hat{H}_{a,c} = \hat{\mathbb{g}}_{a,c} d_{a,c}^{-{\alpha_N} / 2}$ indicate the non-line-of-sight (NLoS) and LoS channel coefficients, respectively. $\alpha_N$ denotes the path loss exponent for NLoS channels, $|\bar{\mathbb{g}}| = 1$ is the LoS channel coefficient, $d_{a,c}$ presents the distance between Alice and Carol, $\hat{\mathbb{g}}_{a,c}$ follows zero-mean unit-variance Gaussian distribution. The links $h_{a,w}$ between Alice and Willie, as well as $h_{a,f}$ between Alice and F-RIS, follow the same Rician fading model as in \eqref{rician}, but the LoS component between Alice and F-RIS differs from that of the other links as $\bar{H}_{a,f} = \bar{\mathbb{g}}_{a,f} d_{a,f}^{-{\alpha_L}/2}$, where $\bar{\mathbb{g}}_{a,f}$ can be expressed as \cite{9508416}
\begin{equation}\label{RIS_Rician}
\begin{aligned}
   \bar{\mathbb{g}}_{a,f} = \sqrt{\gamma_0} [1,e^{-j \pi \sin \iota_{a,f}},...,e^{-j \pi (M-1) \sin \iota_{a,f}}]^T,
\end{aligned}
\end{equation}
where $\iota_{a,f}$ is the incident angle for the transmissions between Alice and F-RIS. The channels $h_{b,f}$ between Bob and F-RIS, $h_{w,f}$ between Willie and F-RIS, and $h_{c,f}$ between Carol and F-RIS, also follows Rician fading as in \eqref{rician} and \eqref{RIS_Rician}.

To enhance the spectrum efficiency in covert transmissions, we apply NOMA for the transmissions between Alice to Bob, and Alice to Carol. Thus, we can have the signal at Carol as
\begin{align}
y_c =~& \sqrt{\beta P_a}\left(h_{a,c} + \boldsymbol{h}^H_{f,c}\boldsymbol{\Theta}\boldsymbol{h}_{a,f}\right)x_c \nonumber \\
&+ \sqrt{(1-\beta) P_a}\left(h_{a,c} + \boldsymbol{h}^H_{f,c}\boldsymbol{\Theta}\boldsymbol{h}_{a,f}\right)x_b + n_c,
\end{align}
where $\beta$ denotes the power allocation factor in NOMA, $\boldsymbol{\Theta} = {\rm diag}(\delta_1 e^{j\theta_1},\delta_2 e^{j\theta_2},...,\delta_M e^{j\theta_M})$ presents the diagonal reflection matrix for F-RIS, with $\delta_{m} \in [0,1]$ and $\theta_m \in [0,2\pi)$ indicate the amplitude and phase shift of $m$th F-RIS element, respectively. $n_c$ denotes the additive white Gaussian noise (AWGN) at Carol. In this work, we applied practical discrete phase shifts as in \cite{Practical}. The relationships between the amplitude, phase shift and incident angle will be described in the next section. Therefore, the signal at Bob can be expressed as
\begin{align}
y_b =~& \sqrt{(1-\beta) P_a}\left(h_{a,b} + \boldsymbol{h}^H_{f,b}\boldsymbol{\Theta}\boldsymbol{h}_{a,f}\right)x_b \nonumber \\
&+ \sqrt{\beta P_a}\left(h_{a,b} + \boldsymbol{h}^H_{f,b}\boldsymbol{\Theta}\boldsymbol{h}_{a,f}\right)x_c + n_b,
\end{align}
where $n_b$ denotes the AWGN at Bob. Considering the principle of successive interference cancelation (SIC) in NOMA \cite{6868214}, if Carol's signal is decoded first and then we decode Bob's signal, the channel rate between Alice and Carol is
\begin{align}
R_c = \log_2\left(1+\frac{\beta P_a|h_{c}|^2}{(1-\beta)P_a|h_{c}|^2+\sigma^2}\right),
\end{align}
where $h_{c} = h_{a,c} + \boldsymbol{h}^H_{f,c}\boldsymbol{\Theta}\boldsymbol{h}_{a,f}$. The achievable rate at Bob is given by
\begin{align}
R_b=\log_2\left(1+\frac{(1-\beta)P_a|h_{b}|^2}{\sigma^2}\right),
\end{align}
where $h_{b} = h_{a,b} + \boldsymbol{h}^H_{f,b}\boldsymbol{\Theta}\boldsymbol{h}_{a,f}$. Notice that since both signals need to be decoded using SIC scheme, it is also possible to first decode Bob's signal and then decode Carol's signal in NOMA transmissions. Then, we can have the received signal at Willie as
\begin{equation}\small\label{willie_signal}
y_{w} = \bigg\{
\begin{array}{ll}
     \sqrt{\beta P_a} h_{w} x_c + n_w, &\mathcal{H}_0\\
     \sqrt{(1-\beta) P_a} h_{w} x_b + \sqrt{\beta P_a} h_{w} x_c + n_w, &\mathcal{H}_1, \\
\end{array}
\end{equation}
where $h_{w} = h_{a,w} + \boldsymbol{h}^H_{f,w}\boldsymbol{\Theta}\boldsymbol{h}_{a,f}$, $n_w$ is the AWGN at Willie. $\mathcal{H}_0$ denotes that Willie thinks Alice is transmitting public signals only, $\mathcal{H}_1$ presents that Willie has detected the covert transmissions.

Willie's optimal detector is based on radio-meter, and the decision rule can be expressed as \cite{10680088}
\begin{align}
  {\mathbb H}(y_w)\underset{D_0}{\overset{D_1}{\gtrless}}\tau,
\end{align}
where $\tau$ indicates the detection threshold in covert communications, $D_0$ and $D_1$ denote Willie's detection results. Thus, the detection error probability at Willie is defined as the sum of false alarm and missed detection probabilities as \cite{10752906}
\begin{align}
  \xi = \mathcal{P}_{FA} + \mathcal{P}_{MD},
\end{align}
where
\begin{align}
  \mathcal{P}_{FA} &= \mathcal{P}(D_1|H_0) = \mathcal{P}({\mathbb H}(y_w)>\tau|\mathcal{H}_0),\\
  \mathcal{P}_{MD} &= \mathcal{P}(D_0|H_1) = \mathcal{P}({\mathbb H}(y_w)\leq\tau|\mathcal{H}_1).
\end{align}

Covert communication requires the detection error probability $\xi$ to satisfy the covert constraint as
\begin{align}
  \xi \geq 1 - \epsilon,
\end{align}
where $\epsilon \in [0,1]$ denotes the threshold that represents the security level in covert communications.

By using Pinsker’s inequality \cite{edwards2008elements}, this constraint can be approximated via Kullback-Leibler (KL) divergence to simplify analysis and optimization. Specifically, the covert constraint can be approximated by the following inequalities involving KL divergence between the probability distributions under hypotheses $H_0$ and $H_1$:
\begin{align}
  D(p_0||p_1) &\leq 2\epsilon^2,\\
  D(p_1||p_0) &\leq 2\epsilon^2,
\end{align}
where $p_0(y_w)$ and $p_1(y_w)$ denote the probability density functions of $y_w$ under $H_0$ and $H_1$, respectively. The KL divergence expressions are given by \cite{10643612}
\begin{align}
  D(p_0||p_1) &= \ln\frac{\lambda_1}{\lambda_0}+\frac{\lambda_0}{\lambda_1}-1,\\
  D(p_1||p_0) &= \ln\frac{\lambda_0}{\lambda_1}+\frac{\lambda_1}{\lambda_0}-1,
\end{align}
where $\lambda_0 = \beta P_a |h_w|^2 + \sigma^2$, $\lambda_1 = P_a|h_w|^2 + \sigma^2$. Therefore, the covert constraints for covert communications can be written as
\begin{align}\small\label{eq:convertConstraint}
\bold{\rm (C1)}: \ln\frac{\lambda_1}{\lambda_0}+\frac{\lambda_0}{\lambda_1}-1\leq 2\epsilon^2,\quad
\ln\frac{\lambda_0}{\lambda_1}+\frac{\lambda_1}{\lambda_0}-1\leq 2\epsilon^2.
\end{align}

These inequalities represent covert communication constraints, which ensure that the probability of Willie successfully distinguishing between $H_0$ and $H_1$ remains sufficiently low to achieve the desired covertness.

\section{F-RIS Model and Problem Formulation} \label{sec:FRIS}
\subsection{F-RIS Model}

\begin{figure}
    \centerline{\includegraphics[width=3.2 in]{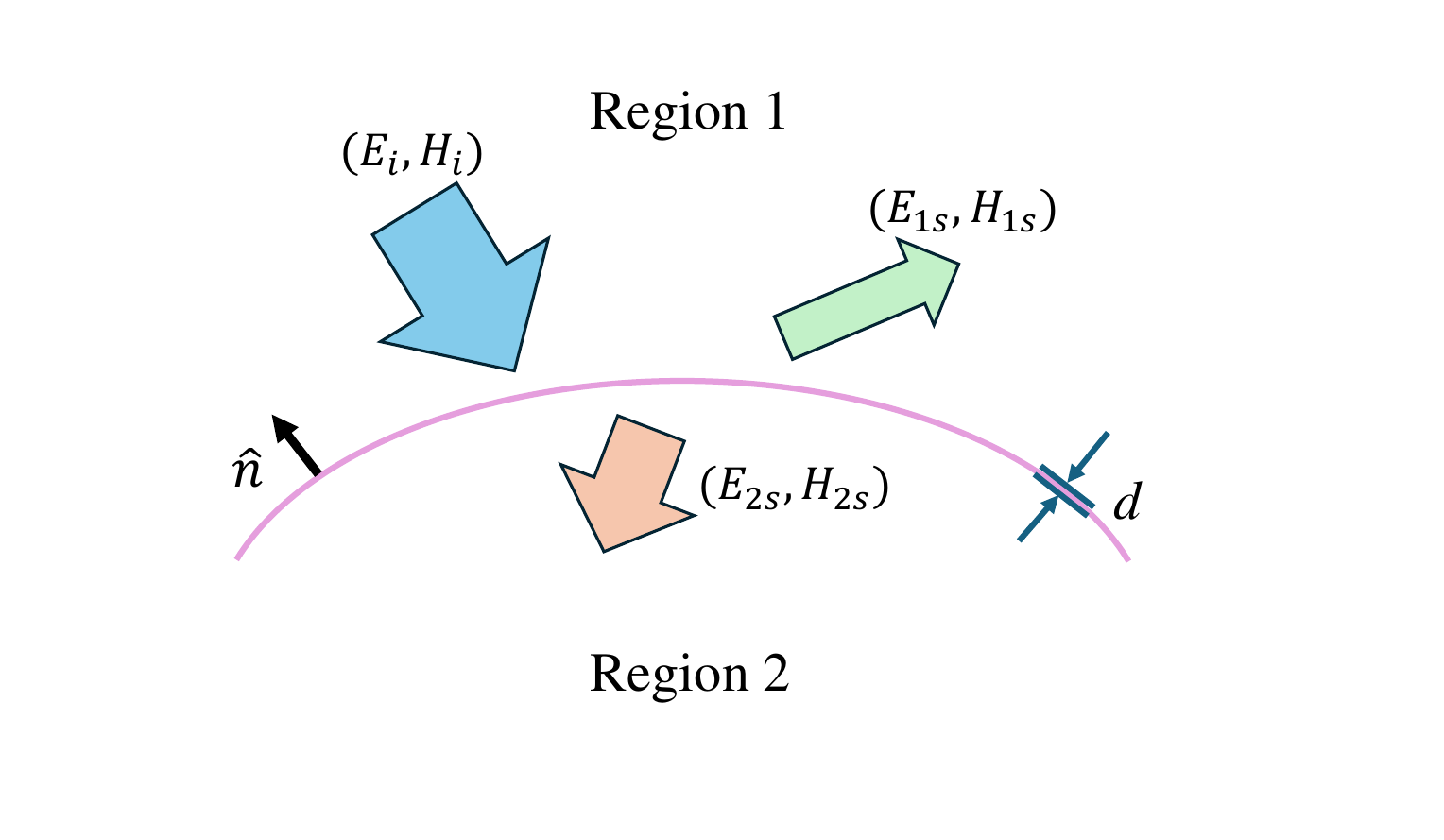}}
    \caption{Transformation of an arbitrary incident wavefront into a desired scattering profile using a transversely inhomogeneous metasurface.}
    \label{WholeSurf}
\end{figure}

The investigation of F-RIS begins with the electromagnetic (EM) characteristics of the overall surface. Fig. \ref{WholeSurf} shows EM wave interactions involving a metasurface whose transverse properties vary arbitrarily. The structure is presumed to be electrically thin, such that its thickness \( d \) is much smaller than the free-space wavelength \( \lambda_0 = 2\pi/k_0 \), where \( k_0 \) denotes the vacuum wave number. Due to this subwavelength dimensionality, the metasurface can be effectively modeled through spatially dependent electric and magnetic surface currents, \( \mathbf{J}_s \) and \( \mathbf{M}_s \), respectively. These currents are governed locally by electric admittance and magnetic impedance tensors: $\mathbf{J}_s = \underline{\underline{\mathbf{Y}}}_e(\mathbf{r}) \cdot \mathbf{E}_t$, $\mathbf{M}_s = \underline{\underline{\mathbf{Z}}}_m(\mathbf{r}) \cdot \mathbf{H}_t,$ where \( \mathbf{E}_t \) and \( \mathbf{H}_t \) denote the tangential components of the EM fields on the surface~\cite{metafilm}.

For the sake of analytical tractability, and considering the problem’s inherent symmetries, the admittance and impedance are treated as scalar quantities \( Y_e \) and \( Z_m \) under the specific excitation conditions considered. The metasurface response may alternatively be described via spatially varying reflection and transmission coefficients, defined as $R = r(\mathbf{r})e^{j\phi_r(\mathbf{r})}$, $T = t(\mathbf{r})e^{j\phi_t(\mathbf{r})}$, corresponding to a locally homogeneous surface excited under normal incidence. According to
\begin{align}
R &= -\frac{2\left( \eta_0^2\, Y_e - Z_m \right)}
           {\left(2 + \eta_0\, Y_e\right)
            \left(2\eta_0 + Z_m\right)}, \\
T &= -\frac{-2 + \eta_0\, Y_e}
          {2 + \eta_0\, Y_e}
    + \frac{2\left( \eta_0^2\, Y_e - Z_m \right)}
           {\left(2 + \eta_0\, Y_e\right)
            \left(2\eta_0 + Z_m\right)},
\end{align}
these coefficients maintain a direct correspondence with the surface’s EM parameters \( Y_e \) and \( Z_m \)~\cite{Wavefront, metafilm2}, with \( \phi_r \) and \( \phi_t \) representing the phases introduced upon reflection and transmission, respectively. $\eta_0$ is the wave impedance of EM waves in free space, which is 377 $\Omega$. In the case of a passive and non-dissipative metasurface, the local power conservation requirement implies that $r^2 + t^2 = 1$, which further necessitates that both \( {Y}_e \) and \( {Z}_m \) be purely imaginary functions.

To achieve the transformation of an arbitrary incident wavefront \((\mathbf{E}_i, \mathbf{H}_i)\) into scattered waves \((\mathbf{E}_{1s}, \mathbf{H}_{1s})\) in region 1 and \((\mathbf{E}_{2s}, \mathbf{H}_{2s})\) in region 2 using a single ultrathin metasurface, the average induced current distributions on the metasurface must be meticulously engineered to compensate for the field discontinuities across the interface. Furthermore, the surface admittance and impedance must be tailored to satisfy the boundary conditions at every point on the metasurface. The following equations represent the precise isotropic metasurface boundary condition required to transform the impinging wavefront into the specified reflected and transmitted waves \cite{metafilm, Boundary1, Boundary2}.
\begin{align}
\hat{n} \times (\mathbf{H}_2 - \mathbf{H}_1)\big|_S &= \frac{1}{2} Y_e (\mathbf{E}_{2t} + \mathbf{E}_{1t})\big|_S, \\
\hat{n} \times (\mathbf{E}_2 - \mathbf{E}_1)\big|_S &= -\frac{1}{2} Z_m (\mathbf{H}_{2t} + \mathbf{H}_{1t})\big|_S,
\end{align}
where $t$ denotes the tangential components of the field in each region. The analysis can be further extended to the anisotropic case to account for polarization coupling.

In the particular case of reflective metasurface, we assume that the incident and scattered fields are linearly polarized transverse-magnetic (TM) plane waves with wave vectors
\(\mathbf{k}_i = k_0 [-\sin(\iota_i)\hat{\mathbf{x}} + \cos(\iota_i)\hat{\mathbf{z}}]\)
and
\(\mathbf{k}_r = -k_0 [\sin(\iota_r)\hat{\mathbf{x}} + \cos(\iota_r)\hat{\mathbf{z}}]\). Consequently, the electric surface admittance and magnetic surface impedance necessary to achieve this scattering signature are given by
\begin{align}
Y_e &= 2 \, \frac{\hat{\mathbf{x}} \cdot (\mathbf{H}_i + \mathbf{H}_{1s})} {\hat{\mathbf{y}} \cdot (\mathbf{E}_i + \mathbf{E}_{1s})} \bigg|_{z=0} \notag \\
&= \frac{2}{\eta_0}
\frac{ e^{-j k_0 \sin(\iota_i) x} - A_r e^{-j k_0 \sin(\iota_r) x}}
{\cos(\iota_i) e^{-j k_0 \sin(\iota_i) x} + A_r \cos(\iota_r) e^{-j k_0 \sin(\iota_r) x}}, \\
Z_m &= 2 \, \frac{\hat{\mathbf{y}} \cdot (\mathbf{E}_i + \mathbf{E}_{1s})} { \hat{\mathbf{x}} \cdot (\mathbf{H}_i + \mathbf{H}_{1s})} \bigg|_{z=0} \notag \\
&= 2 \eta_0
\frac{\cos(\iota_i) e^{-j k_0 \sin(\iota_i) x} + A_r \cos(\iota_r) e^{-j k_0 \sin(\iota_r) x}}
{e^{-j k_0 \sin(\iota_i) x} - A_r e^{-j k_0 \sin(\iota_r) x}},
\end{align}
where $A_r$ represents the normalized amplitude of the electric field in the reflected plane wave, relative to the incident field. Significantly, $|A_r| = \sqrt{\cos \iota_i}/\sqrt{{\cos \iota_r}}$ ensures that the power reflected along the direction normal to the surface matches the impinging power. Although the required overall surface impedance conditions for improving reflection performance are known, the variability of F-RIS limits the analysis using this approach.

Therefore, we propose to divide the surface into individual units for a more efficient analysis of its reflection characteristics due to the varying degrees of bending across different regions. Given that the metasurface-based reflection units have subwavelength structures, when a plane wave is incident on F-RIS, the process can be viewed as the plane wave interacting with each subwavelength unit at different angles. Consequently, the reflection characteristics are examined from the perspective of the individual reflection units.

Typically, as the incident angle of the wave changes, both the amplitude and phase of the reflection characteristics of the RIS are modified. Changing the incident angle from normal to oblique alters the ratio of the transverse components of the electric and magnetic fields on the surface, which, in turn, affects the wave impedance. This results in an increasing impedance mismatch for steeper angles.

\begin{figure}
    \centerline{\includegraphics[width=3 in]{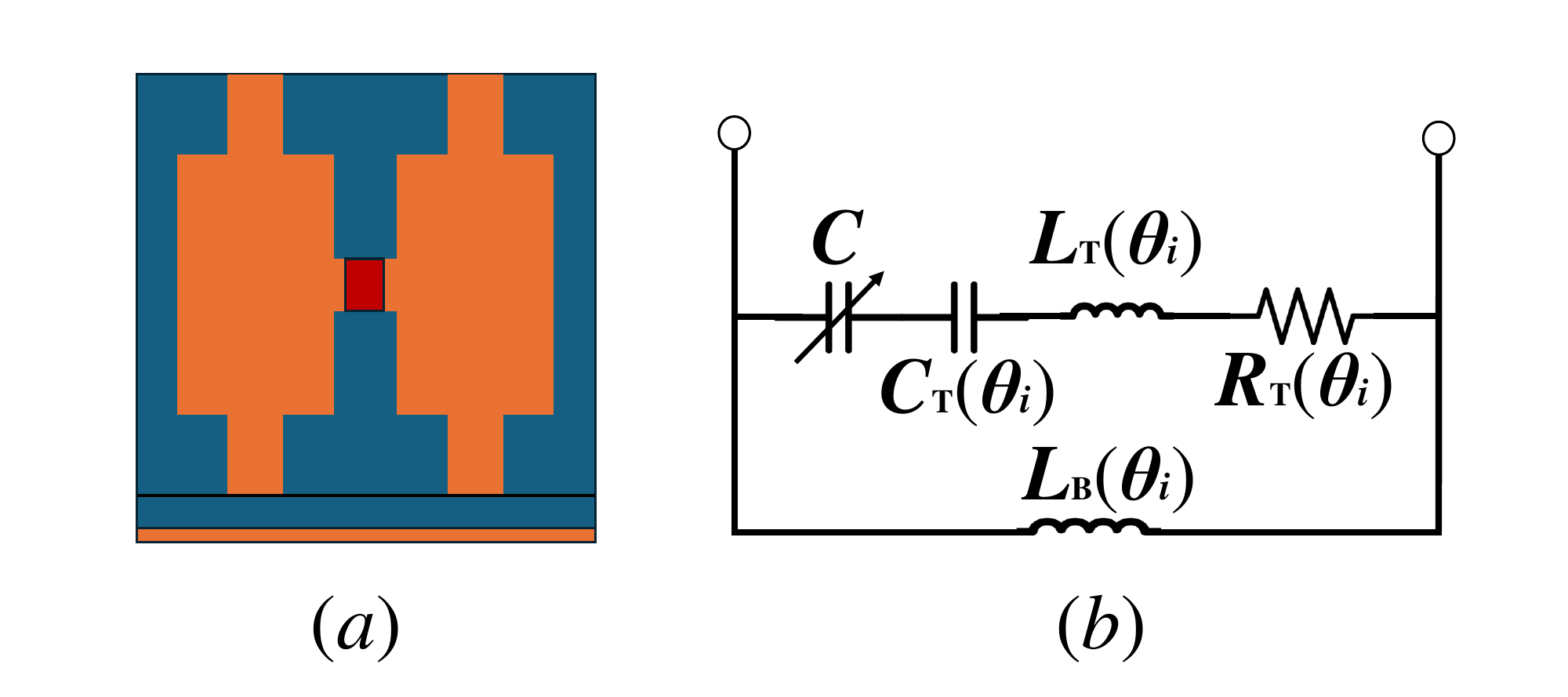}}
    \caption{The schematic of the unit cell of the proposed RIS in \cite{2bit} and its equivalent circuit model. }
    \label{2BitUC}
\end{figure}

To verify the influence of the incident angle on the reflection performance, an angle-dependent phase shifter model was proposed in \cite{2bit}. The structure of the unit cell and its corresponding equivalent circuit model are shown in Fig. \ref{2BitUC}. $C$ is the variable capacitance of the varactor, $R_T(\iota_i)$, $L_T(\iota_i)$, and $C_T(\iota_i)$ are the equivalent resistance, inductance and capacitance of the top layer. $L_B(\iota_i)$ represents the effective inductance caused by the substrate and the bottom layer. The dielectric and metal ground can be regarded as a short-circuited transmission line model, and its input impedance at normal incidence can be expressed as:
\begin{equation}
Z_{B} =j\frac {\eta _{0}}{\sqrt {\varepsilon _{r}}}\tan \frac {2\pi f\sqrt {\varepsilon _{r}} H}{c},
\end{equation}
where $H$ is the thickness of the substrate, and $\epsilon_r$ denotes the relative permittivity of the substrate. Since $Z_B$ is purely imaginary, it can be interpreted as an inductance.

According to the equivalent circuit model, the impedance of the unit cell can be written as
\begin{align}
&Z ({\iota_i,\omega,C}) \\=&\frac {{j\omega{L_{\mathrm{ B}}}(\iota_i)\left ({{{R_{\mathrm{ T}}}(\iota_i) + j\omega{L_{\mathrm{ T}}}(\iota_i) + \frac {1}{{ {j\omega{C_{\mathrm{ T}}}(\iota_i)} }} + \frac {1}{j\omega C}} }\right)}}{{j\omega{L_{\mathrm{ B}}}(\iota_i) + \left ({{{R_{\mathrm{ T}}}(\iota_i) + j\omega{L_{\mathrm{ T}}}(\iota_i) + \frac {1}{{ {j\omega{C_{\mathrm{ T}}}(\iota_i)} }} + \frac {1}{j\omega C}} }\right)}}.
\end{align}
From this equation, it is known that the impedance of the unit cell is related to the incident angle $\iota_i$, the operating frequency $f$, and the capacitance of the varactor $C$. Consequently, the reflection coefficient can be calculated as:
\begin{equation}
\Gamma =\frac {Z ({\iota_i,\omega,C}) -\eta_0}{Z ({\iota_i,\omega,C}) +\eta_0}
\end{equation}.

Considering the parameters adopted in \cite{2bit} and slight adjustments, the parameters we use at normal incidence are as follows: $L_B(\iota_i)$ = 15.83 nH, $L_T(\iota_i)$ = 38.26 nH, $R_T(\iota_i)$ = 2.2 $\Omega$, $C_T(\iota_i)$ = 15.6 pF, $C$ varies from 0.63 to 2.67 pF. Following the simulation results in \cite{2bit}, we fitted the relationship between the reflection amplitude $\delta$, the reflection phase $\theta$, and the incident angle $\iota$ for angles within 45°. The resulting fitted function is as follows:
\begin{equation}
\begin{split}
\delta(\theta, \iota_i) = & \, p1 + p2 \cdot \sin(\theta) - p3 \cdot \cos(\theta) \\
& + p4 \cdot \iota_i + p5 \cdot \iota^2_i \\
& + p6 \cdot \sin(\theta) \cdot \iota_i + p7 \cdot \cos(\theta) \cdot \iota_i,
\end{split}
\end{equation}
where array $\boldsymbol{\mathit{p}} = \{p1, p2, p3, p4, p5, p6, p7\}$ represents the weighting parameters obtained through curve fitting based on actual measurements for F-RIS. The fitting equation is formulated using a trigonometric–polynomial hybrid expression. The choice of basis functions is motivated by the underlying physics of wave reflection. Specifically, the dependence of the reflection amplitude on the reflection phase is inherently oscillatory. Therefore, a combination of $\sin(\theta)$ and $\cos(\theta)$ terms is introduced to capture the periodic modulation. In contrast, the incidence angle contributes a non-periodic variation that governs the overall envelope of the reflection response, which is approximated by a low-order polynomial in $\iota_i$ to represent both linear and weakly nonlinear angular trends. Furthermore, interaction terms such as $\sin(\theta) \cdot \iota_i$ and $\cos(\theta) \cdot \iota_i$ are included to account for the coupling between phase oscillations and angle-dependent modulation of their amplitudes. Finally, a constant term defines the baseline reflection level.

\begin{figure}[t!]
  \centering
  \centerline{\includegraphics[scale=0.062]{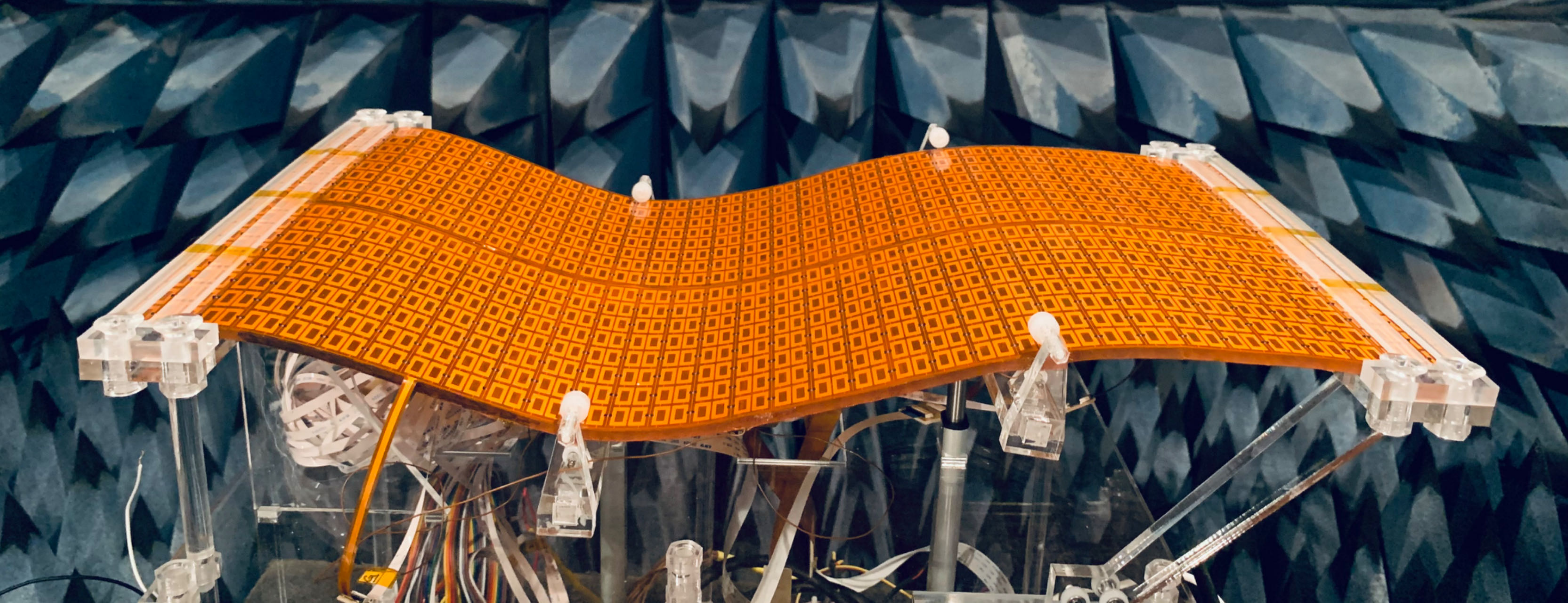}}
 \caption{\small Illustration of F-RIS.} \label{fig:FRIS1}
\end{figure}

However, as the incident angle increases further, the reflection characteristics of conventional reflective RIS degrade significantly, greatly limiting the applicability of F-RIS. It is worth noting that, within the framework of this model, the maximum bending angle is constrained to $90^{\circ}$. To address this, some studies on reflective RIS focus on angle insensitivity. In \cite{3bit}, a phase range of up to 315° is achieved, with eight states (3 bits), each separated by a stable interval of 45°, effectively accommodating a wide range of incidence angles from 0° to 60°.

The primary function of metallic vias is to mitigate the impact of large incident angles on the reflection characteristics of the RIS, thereby ensuring phase stability within a certain range of incident angles. This structural design partially decouples the relationship between reflection properties and the angle of incidence, offering greater adaptability for F-RIS implementations. However, metallic vias inherently constrain the flexibility of the structure. As a solution, flexible substitutes could be employed as in Fig. \ref{fig:FRIS1} \cite{li2025flexible}. 



\subsection{Problem Formulation}
In this work, we consider the F-RIS-aided covert transmissions between UAVs, with public transmission between Alice and Carol. Thus, we aim to maximize the covert transmission rate at Bob by jointly optimizing UAV trajectories, power allocation, F-RIS reflection vectors and incident angle. The proposed problem formulation is expressed as
\begin{align}
    &\max_{Q(t), \beta(t), \vartheta(t), \iota(t)} \frac{1}{T}\sum_{t=1}^{T} R_b(t),\label{SecrecyFunc}\\
    {\rm s.t.}&~ \frac{1}{T}\sum_{t=1}^{T} R_c(t) \geq \varepsilon_c \tag{\ref{SecrecyFunc}{a}}, \label{SecrecyFuncSuba}\\
    &\delta_{m} \in [0,1], \theta_m \in [0,2\pi) \tag{\ref{SecrecyFunc}{b}}, \label{SecrecyFuncSubb}\\
    &v_a(t) \leq v_{\rm max}, v_b(t) \leq v_{\rm max} \tag{\ref{SecrecyFunc}{c}}, \label{SecrecyFuncSubc}\\
    &q^a_z(t) \in [q_{z,\rm min},q_{z,\rm max}], q^b_z(t) \in [q_{z,\rm min},q_{z,\rm max}] \tag{\ref{SecrecyFunc}{d}}, \label{SecrecyFuncSubd}\\
    &ac_a(t) \in [0, ac_{\rm max}], ac_b(t) \in [0, ac_{\rm max}] \tag{\ref{SecrecyFunc}{e}}, \label{SecrecyFuncSube}\\
    &\zeta^h_a(t) \in [0,2\pi), \zeta^h_b(t) \in [0,2\pi) \tag{\ref{SecrecyFunc}{f}}, \label{SecrecyFuncSubf}\\
    &\zeta^p_a(t) \in [0,\pi), \zeta^p_b(t) \in [0,\pi) \tag{\ref{SecrecyFunc}{g}}, \label{SecrecyFuncSubg}\\
    &\beta \in (0, 1) \tag{\ref{SecrecyFunc}{h}}, \label{SecrecyFuncSubh}\\
    &\bold{\rm (C1)} \tag{\ref{SecrecyFunc}{i}}, \label{SecrecyFuncSubi}\\
    &d_{a,b} \geq d_{\rm min}, \tag{\ref{SecrecyFunc}{j}}, \label{SecrecyFuncSubj}
\end{align}
where $T$ is the number of total time slots, $Q(t) = \{q_a(t), q_b(t)\}$ is the UAV trajectories, $\vartheta(t) = {\theta_1, ..., \theta_M}$ denotes the phase shifts for F-RIS, $\iota(t) = \{\iota_h(t), \iota_p(t)\}$ denotes the change in the relative angle with respect to the original position due to the shape deformation of elements within F-RIS, $\iota_h(t) = \{\iota_{h,1}, ..., \iota_{h,M}\} $ denotes the rotation angle with respect to the horizontal plane, $\iota_v(t) = \{\iota_{v,1}, ..., \iota_{v,M}\}$ denotes the rotation angle with respect to the vertical plane. \eqref{SecrecyFuncSuba} indicates that the average communication rate of public transmissions should meet the threshold $\varepsilon_c$. \eqref{SecrecyFuncSubb} presents the physical constraints of amplitudes and phase shifts. \eqref{SecrecyFuncSubc} shows the maximum speed limitation $v_{\rm max}$ for UAVs. \eqref{SecrecyFuncSubd} indicates the minimum altitude $q_{z,\rm min}$ and maximum altitude $q_{z,\rm max}$ for UAVs. \eqref{SecrecyFuncSube} introduces the maximum acceleration $ac_{\rm max}$ for UAVs. \eqref{SecrecyFuncSubf} and \eqref{SecrecyFuncSubg} describe the limitation of heading angle and the pitch angle for UAVs, respectively. \eqref{SecrecyFuncSubh} is the range of NOMA power allocation factor. \eqref{SecrecyFuncSubi} denotes the covert constraints in the proposed networks. \eqref{SecrecyFuncSubj} shows that the distance between UAVs need to be larger than the threshold $d_{\rm min}$ for safety flying. Considering the time-varying Rician fading channels and the dynamic positions of UAVs, this makes \eqref{SecrecyFunc} as a long-term optimization problem. To address this problem in a highly dynamic UAV network, we employ a DRL algorithm to learn a strategy aimed at optimizing the average covert transmission rate while satisfying constraints in \eqref{SecrecyFunc}. This approach offers an adaptive solution for dynamic UAV networks and highlights the promising potential of F-RIS in covert communications.

\section{DRL-Based Optimization} \label{sec:DRL}
To effectively address the complicated optimization problem in \eqref{SecrecyFunc}, we need to model the system as a Markov Decision Process (MDP) first. The primary components of the MDP in our scenario are state space, action space and reward function. The state is used to describe the varying parameters in the system, we assume at time slot $t$ the state is
\begin{align}
s(t) = \{t, {\mathbb H}(t), Q(t), \frac{1}{t}\sum_{i=1}^{t} R_b(i), \frac{1}{t}\sum_{i=1}^{t} R_c(i)\},
\end{align}
where ${\mathbb H}(t)$ includes all the channel coefficients at time slot $t$. Action is utilized to indicate the optimization variables in \eqref{SecrecyFunc}, so we assume action as
\begin{equation}
\begin{aligned}
a(t) =~& \{ac_a(t), ac_b(t), \zeta^h_a(t), \zeta^h_b(t), \zeta^p_a(t), \zeta^p_b(t),\\
&\beta(t), \vartheta(t), \iota(t)\}.
\end{aligned}
\end{equation}
Reward function is used to provide feedbacks to the agent in DRL to help learn a policy for achieving the goal of the optimization problem. We design the reward function as
\begin{align}
r(t) = \frac{1}{t}\sum_{i=1}^{t} R_c(i) + \nu_1 \mu\bigg(\frac{1}{t}\sum_{i=1}^{t} R_c(i) \geq \varepsilon_c \bigg) + \nu_2 \mu\bigg(\rm (C1)\bigg),
\end{align}
where $\nu_1$ and $\nu_2$ are weights for constraints, $\mu(\cdot) = 0$ when $(\cdot)$ is true, otherwise $\mu(\cdot) = -1$. The weight factors $\nu_1$ and $\nu_2$ are flexible parameters that can be dynamically adjusted to reflect different transmission requirements or system priorities. Considering the complexity and dynamic features of the proposed F-RIS-aided UAV network, we employ the distributed soft actor-critic (DSAC) method \cite{10745225} which integrates a distributional perspective to capture reward uncertainties more accurately.

The soft return distribution is first updated as
\begin{equation}
\Re_\pi(s(t),a(t)) = r(t)+\varpi \varrho(t+1),
\end{equation}
where policy $\pi$, discount factor $\varpi$, and the cumulative return $\varrho(t)$ is defined as
\begin{equation}
\varrho(t) = \sum_{i=t}^{\infty}\varpi^{(i-t)}[r(i)-\alpha_e \log\pi(a(i)|s(i))],
\end{equation}
where $\alpha_e$ indicates the entropy regularization parameter in DSAC.

The state-action soft value is thus calculated as
\begin{equation}
J_{\pi}(s(t),a(t)) = \mathbb{E}\left[\Re_{\pi}(s(t),a(t))\right].
\end{equation}

Considering the distributional approach, the soft Bellman distribution operator is then given by
\begin{equation}
\begin{aligned}
\mathcal{T}\Re_{\pi}(s(t),a(t)) = ~&r(t)+\varpi[\Re_{\pi}(s(t+1),a(t+1))\\
&-\alpha_e \log\pi(a(t+1)|s(t+1))].
\end{aligned}
\end{equation}

Consequently, the updated soft return distribution can be found via minimizing distributional divergence
\begin{equation}
\hat{\Re}_{\text{new}} = \arg\min\mathbb{E}\left[d_s(\mathcal{T}\hat{\Re}_{\text{old}}(\cdot|s(t),a(t)),\hat{\Re}(\cdot|s(t),a(t)))\right],
\end{equation}
with the divergence typically measured by Kullback-Leibler divergence.

To optimize the soft return, the loss function in DSAC is introduced as
\begin{equation}
L_{\hat{\Re}}(\varsigma) = -\mathbb{E}_{\Phi_{L_{\hat{\Re}}(\varsigma)}}\left[\log\mathcal{P}(G_{\pi_{\phi'}}(s(t),a(t))|\hat{\Re}(\cdot|s(t),a(t)))\right],
\end{equation}
where $\varsigma$ and $\phi'$ are the parameters in DSAC, sampling process $\Phi_{L_{\hat{\Re}}(\varsigma)}$ is defined in \cite{9448360}. Correspondingly, the gradient update of parameter $\varsigma$ is obtained by
\begin{equation}\label{eq:softupdateTheta}
\begin{aligned}
\nabla_{\varsigma}L_{\hat{\Re}}(\varsigma) = ~&-\mathbb{E}_{\Phi_{\nabla_{\varsigma}L_{\hat{\Re}}(\varsigma)}}[\nabla_{\varsigma}\log\mathcal{P}(G_{\pi_{\phi'}}(s(t),\\
&a(t))|\hat{\Re}(\cdot|s(t),a(t)))],
\end{aligned}
\end{equation}
with detailed sampling definitions following Eq. (24) in the original source.

We further explicitly rewrite the gradient update rule for $\varsigma$ by
\begin{equation}
\begin{aligned}
\nabla_{\varsigma}L_{\hat{\Re}}(\varsigma) = ~&\mathbb{E}_{\Phi_{\nabla_{\varsigma}L_{\hat{\Re}}(\varsigma)}}[-\frac{\nabla \Psi_{\hat{\Re}}(\varsigma)}{\nabla J_{\varsigma}(s(t),a(t))}\frac{\nabla J_{\varsigma}(s(t),a(t))}{\nabla \varsigma}\\
&-\frac{\nabla \Psi_{\hat{\Re}}(\varsigma)}{\nabla\sigma_{\varsigma}(s(t),a(t))}\frac{\nabla\sigma_{\varsigma}(s(t),a(t))}{\nabla\varsigma}],
\end{aligned}
\end{equation}
where
\begin{equation}
\begin{aligned}
\frac{\partial\nabla \Psi_{\hat{\Re}}(\varsigma)}{\nabla J_{\varsigma}(s(t),a(t))} = \frac{\mathcal{T}G_{\pi_{\phi'}}(s(t),a(t))-J_{\varsigma}(s(t),a(t))}{\sigma_{\varsigma}^2(s(t),a(t))},
\end{aligned}
\end{equation}

\begin{equation}
\begin{aligned}
\frac{\nabla \Psi_{\hat{\Re}}(\varsigma)}{\nabla\sigma_{\varsigma}(s(t),a(t))} = ~& \frac{[\mathcal{T}G_{\pi_{\phi'}}(s(t),a(t))-J_{\varsigma}(s(t),a(t))]^2}{\sigma_{\varsigma}^3(s(t),a(t))}\\
&- \frac{1}{\sigma_{\varsigma}(s(t),a(t))}.
\end{aligned}
\end{equation}

To stabilize the training, the standard deviation is bounded by
\begin{equation}
\sigma_{\varsigma}(s(t),a(t))=\max(\sigma_{\varsigma}(s(t),a(t)),\sigma_{\min}),
\end{equation}
or using gradient clipping as
\begin{equation}
\begin{aligned}
\frac{\nabla \Psi_{\hat{\Re}}(\varsigma)}{\nabla\sigma_{\varsigma}(s(t),a(t))} = \frac{s-J_{\varsigma}(s(t),a(t))}{\sigma_{\varsigma}^3(s(t),a(t))}-\frac{1}{\sigma_{\varsigma}(s(t),a(t))},
\end{aligned}
\end{equation}
where
\begin{equation}
\begin{aligned}
s =~&\text{clip}(\mathcal{T}G_{\pi_{\phi'}}(s(t),a(t)),J_{\varsigma}(s(t),a(t))-\epsilon,\\
&J_{\varsigma}(s(t),a(t))+\epsilon).
\end{aligned}
\end{equation}

Finally, the policy improvement steps (Eqs. (31)-(32) original PDF) are defined explicitly as
\begin{equation}
\begin{aligned}
L_{\pi}(\phi) =~& \mathbb{E}_{s(t)\sim\mathcal{B},a(t)\sim\pi_{\phi}}[J_{\varsigma}(s(t),a(t))\\
&-\alpha_e \log(\pi_{\phi}(a(t)|s(t)))],
\end{aligned}
\end{equation}

\begin{equation}
\begin{aligned}
\nabla_{\phi}L_{\pi}(\phi) = ~&\mathbb{E}_{s(t)\sim\mathcal{B},\rho}[-\alpha_e \nabla_{\phi}\log(\pi_{\phi}(a(t)|s(t)))\\
&(\nabla_{a(t)}J_{\varsigma}(s(t),a(t))\\
&-\alpha_e \nabla_{a(t)}\log(\pi_{\phi}(a(t)|s(t))))\nabla_{\phi}\mathfrak{f}(\rho;s(t))].
\end{aligned}
\end{equation}
DSAC directly learns the continuous return distribution by maintaining the variance of the state-action return within a reasonable range, thereby addressing the issues of gradient explosion and vanishing. This makes it highly beneficial for optimizing solutions to complex problems in long-term learning scenarios. As a pre-trained approach, it enables real-time optimization with very low computational overhead. The pseudo-code of the proposed DRL-based algorithm is in Algorithm \ref{Algorithm DSAC}.

\begin{algorithm}[t!]
\caption{\textbf{DSAC-Based Optimization}: }\label{Algorithm DSAC}
\begin{algorithmic}[1]
 \makeatletter\setcounter{ALG@line}{0}\makeatother
 \State Initialize the parameters in F-RIS-aided UAV networks.
 \State Initialize the neural networks of DSAC.
 \Repeat:
 \For {time slot 1, 2,... T}
 \State Select action $a(t)$ based on the current state $s(t)$
 \State Achieve $s(t+1)$ and $r(t)$ based on $s(t)$ and $a(t)$.
 \State Build a sample $\{s(t),a(t),r(t),s(t+1)$.
 \State Save the sample to the replay buffer $\mathbb{U}$.
 \State Update $\varsigma$ based on \eqref{eq:softupdateTheta}.
 \EndFor
 \For {each epoch}
 \State Randomly select a minibatch from buffer $\mathbb{U}$.
 \State Soft update target and critic networks .
 \EndFor
 \Until Convergence.
\end{algorithmic}
\end{algorithm}

\section{Simulation Results} \label{sec:sim}
\begin{figure}[t!]
  \centering
  \centerline{\includegraphics[scale=0.12]{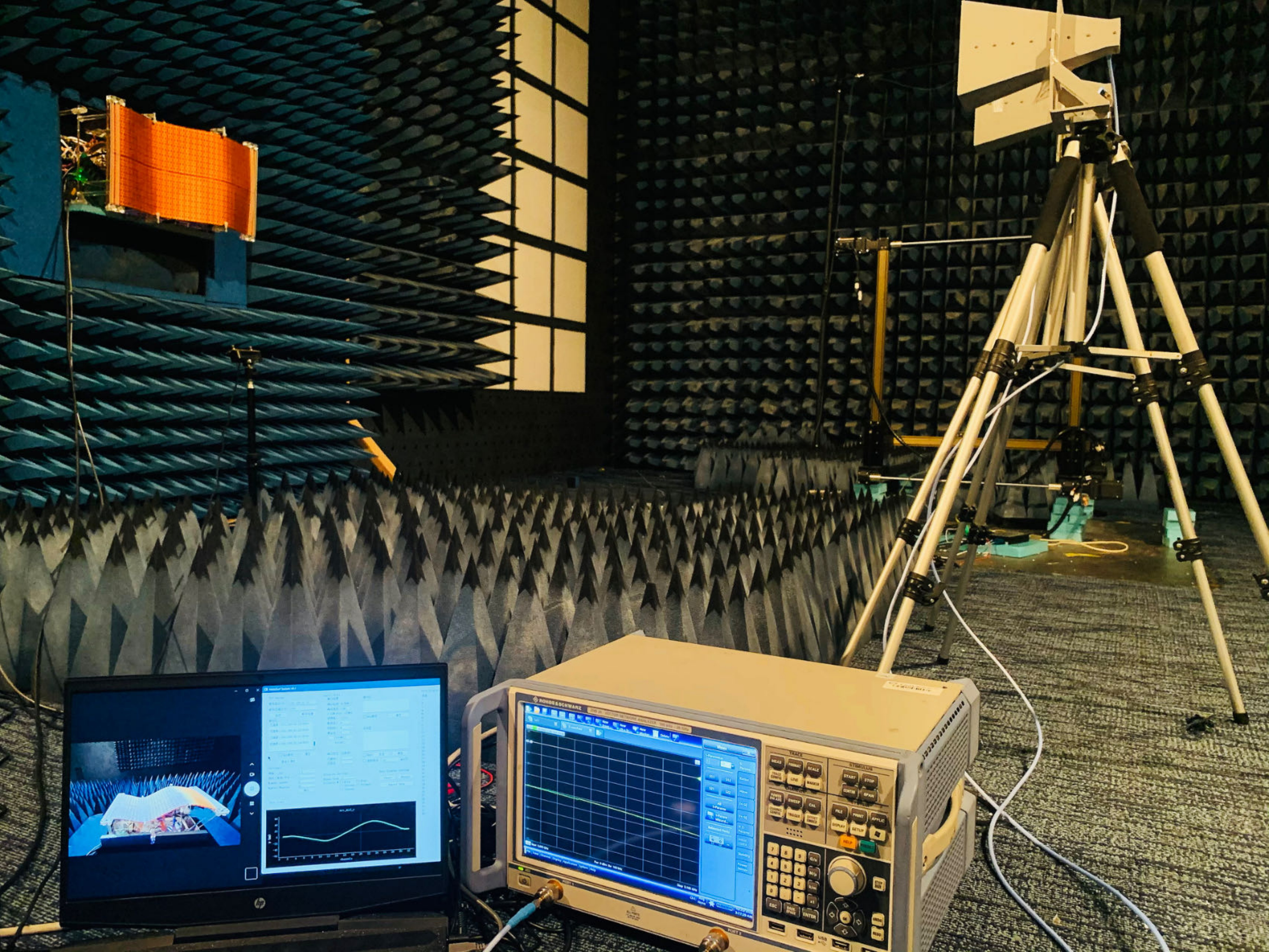}}
 \caption{\small F-RIS experimentation platform.} \label{fig:FRIS}
\end{figure}

Unless otherwise stated, we set simulation parameters as follows: the transmit power $P_a = 0.2$ W, the Rician factor $\kappa = 10$ dB, the path loss exponents of LoS and NLoS channels are 2 and 3, respectively. $\gamma_0 = -10$ dB, the number of reflection elements $M = 64$, covert transmission threshold $\epsilon = 0.1$, public transmission threshold $\varepsilon_c = 3.3$ bps/Hz, $v_{\rm max} = 5$ m, $q_{z,\rm min} = 35$ m, $q_{z,\rm max} = 60$ m, $ac_{\rm max} = 2$ m/s, $d_{\rm min} = 60$ m, the noise level is -50 dBm, $\boldsymbol{\mathit{p}}$ = [0.8816, 0.0473, -0.1010, 0.0004, -0.000019, 0.000055, 0.000321], the discount factor $\varpi = 0.99$. Notice that the weighting parameters of F-RIS are measured from F-RIS experimentation platform. We introduce the multi-agent proximal policy optimization (MAPPO) algorithm as the benchmark in simulation results. We employed F-RIS and UAV testing platform as shown in Fig. \ref{fig:FRIS} to measure the channel coefficients, which were subsequently used for modeling and analysis.

\begin{figure}[t!]
  \centering
  \centerline{\includegraphics[scale=0.6]{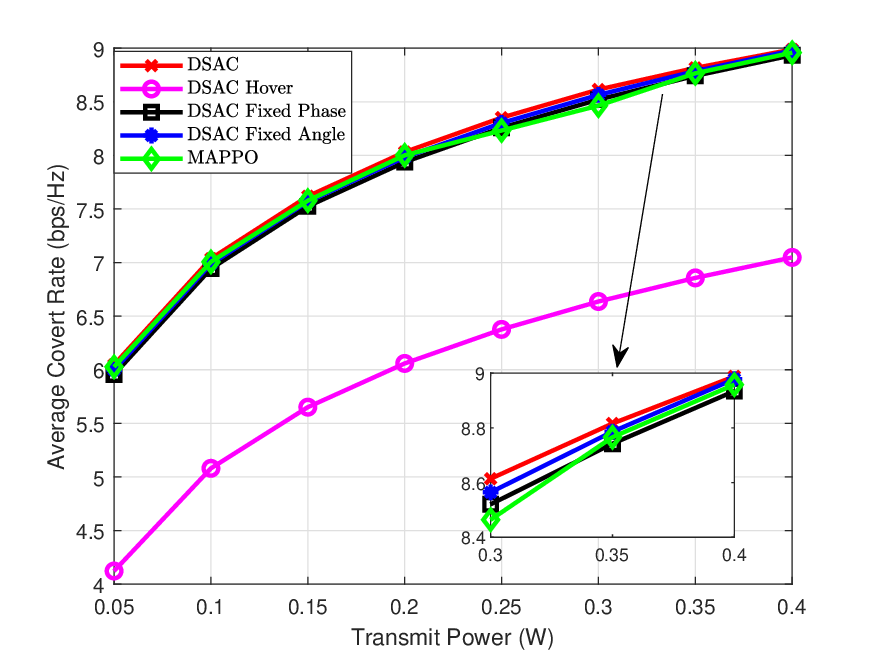}}
 \caption{\small Average covert rate versus the transmit power of Alice.} \label{fig:R1}
\end{figure}

We compare the average covert rate with different transmit power $P_a$ in the proposed system in Fig. \ref{fig:R1}. `DSAC Hover' denotes the optimization excludes UAV trajectory planning, `DSAC Fixed Phase' utilizes fixed phase shifts in optimization, while `DSAC Fixed Angle' uses fixed incident angle which means it is the same as a traditional RIS. It can be observed that the UAV trajectory has the greatest impact on overall performance, as the UAV's high mobility enables rapid changes in link conditions. Moreover, the proposed algorithm outperforms MAPPO regardless of the transmit power. The results from the baseline with fixed phase and fixed incident angle indicate that both factors influence the performance. Currently, the impact of a fixed incident angle is relatively small because, for F-RIS, the effect of varying incident angles has already been minimized during the design phase, enabling near-ideal performance under diverse environmental conditions. Therefore, as the influence of F-RIS deformation on performance becomes increasingly negligible, F-RIS—which can cover objects of arbitrary shapes and freely alter its form—will emerge as a crucial technology in future communication systems.

\begin{figure}[t!]
  \centering
  \centerline{\includegraphics[scale=0.6]{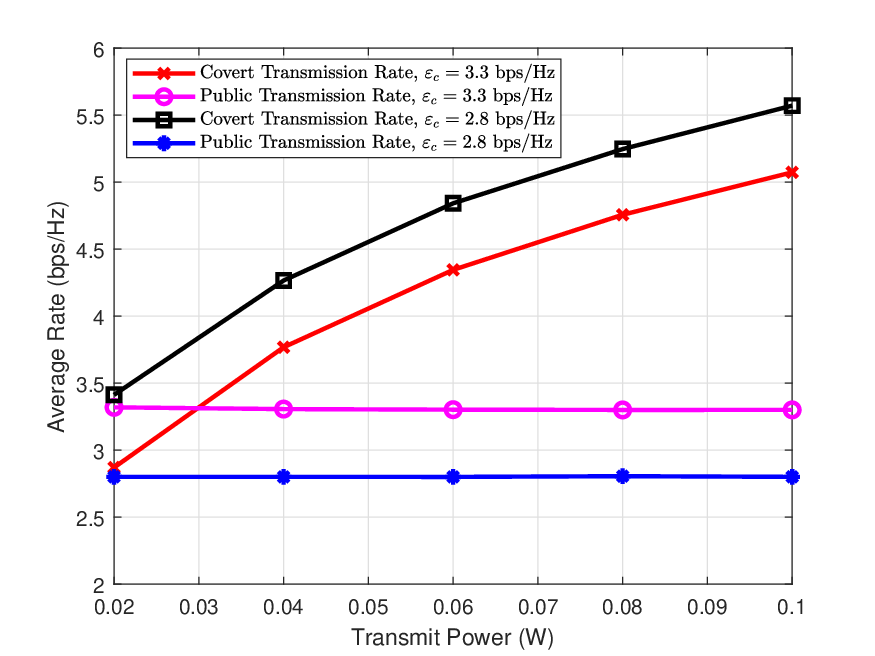}}
 \caption{\small Average covert and public transmission rates versus the transmit power of Alice.} \label{fig:R1_2}
\end{figure}
As shown in Fig. \ref{fig:R1_2}, as the transmit power increases, the signal is enhanced to improve the performance of covert communications. When the threshold of the public transmission is set at 2.8 bps/Hz, Carol's lower transmission requirement allows for allocating more transmit power and reflection signals to Bob, consequently increasing the covert communication rate. Since the public transmission serves only as a constraint in the optimization problem, the DRL-based strategy needs merely to satisfy its minimum threshold requirement, and the remaining resources can be utilized to enhance the covert communication performance. Therefore, Carol's transmission performance consistently matches the threshold value.

\begin{figure}[t!]
  \centering
  \centerline{\includegraphics[scale=0.6]{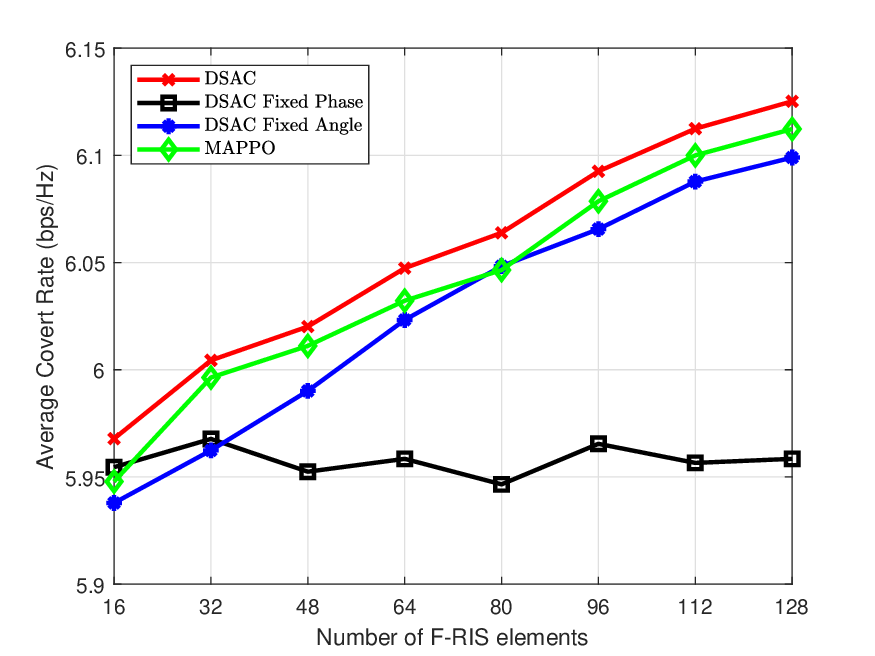}}
 \caption{\small Average covert rate versus the number of F-RIS reflection elements.} \label{fig:R2}
\end{figure}

From Fig. \ref{fig:R2}, we compare the average covert transmission rate with different numbers of F-RIS reflection elements. As illustrated in the figure, the number of elements of F-RIS significantly affects the overall performance. With an increased number of reflecting elements, the carefully optimized reflecting array can better strengthen the relevant channels, enhancing the covert transmission rate between Alice and Bob, while ensuring the quality of public communication between Alice and Carol. It can also be observed from the results that, for schemes without optimized F-RIS phase control, increasing the number of reflecting elements does not yield meaningful performance improvements. Thus, phase optimization of the F-RIS plays a pivotal role in the proposed optimization problem. Furthermore, these results demonstrate the superiority of the proposed algorithm and confirm that the F-RIS possesses the inherent advantage of adaptively modifying the electromagnetic environment.

\begin{figure}[t!]
  \centering
  \centerline{\includegraphics[scale=0.6]{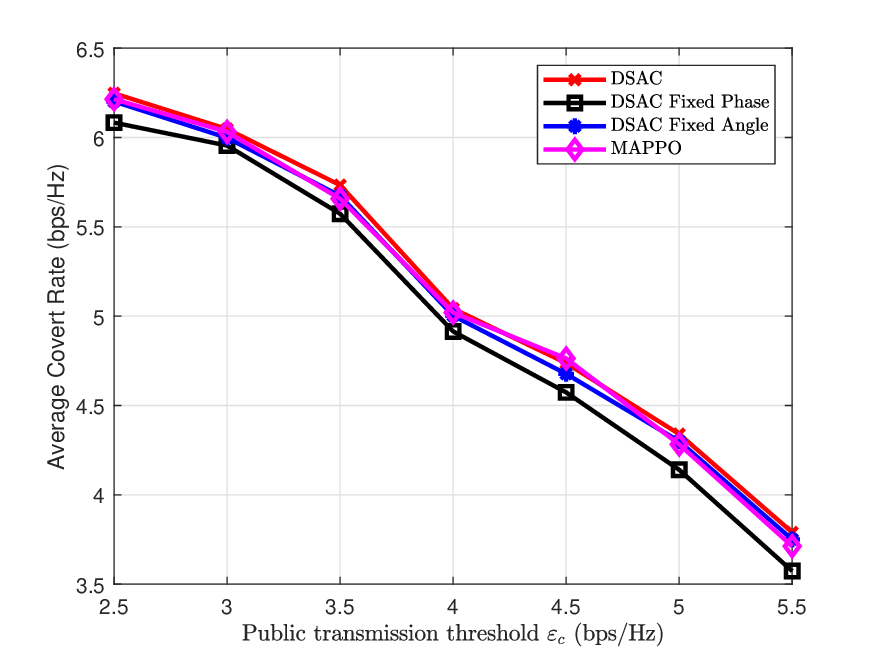}}
 \caption{\small Average covert rate versus the public transmission threshold $\varepsilon_c$.} \label{fig:R3}
\end{figure}

In Fig. \ref{fig:R3}, we compare the average covert rate versus the public transmission threshold $\varepsilon_c$. As observed from the figure, as the required rate for public communication increases, the performance of covert communication correspondingly decreases. This occurs because Alice must simultaneously support both public and covert communications. Even when employing NOMA technology to enhance spectrum utilization, increasing Carol's signal strength inevitably diminishes the transmission capability toward Bob. Moreover, covert communication must satisfy certain secrecy constraints, which further complicates maintaining covert communication performance when the transmitted signal strength is reduced.

\begin{figure}[t!]
  \centering
  \centerline{\includegraphics[scale=0.6]{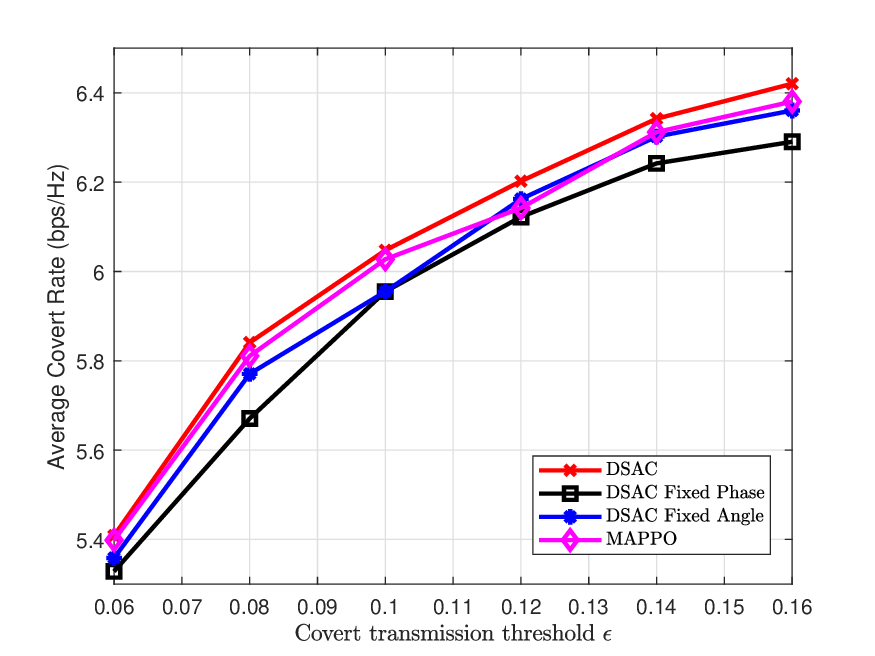}}
 \caption{\small Average covert rate versus the covert transmission threshold $\epsilon$.} \label{fig:R4}
\end{figure}

Considering that the covert constraint plays the key role in covert communications, we demonstrate the relationship between average covert rate and the covert transmission threshold $\epsilon$ in Fig. \ref{fig:R4}. As illustrated in the figure, relaxing the constraints on covert communication results in an increased covert transmission rate. This is because a higher covert transmission threshold implies greater difficulty for Willie to detect the communication between Alice and Bob, thus enabling a safer increase in the covert transmission rate. Furthermore, variations in phase and incident angle significantly impact the performance of covert communication, highlighting the critical importance of F-RIS in this context.

\begin{figure}[t!]
  \centering
  \centerline{\includegraphics[scale=0.6]{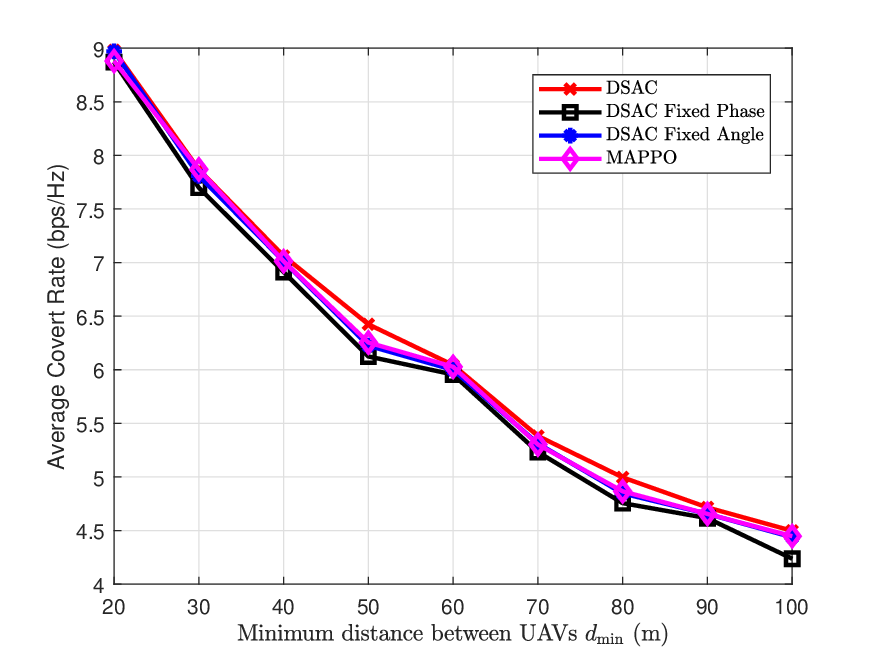}}
 \caption{\small Average covert rate versus the minimum distance between UAVs $d_{\rm min}$.} \label{fig:R5}
\end{figure}

To analyze and compare the proposed method, we conduct an experiment with different minimum distance between UAVs. This figure illustrates that reducing the distance between UAVs significantly enhances their covert communication rate. This is because distance is one of the critical factors causing communication signal loss. As the UAVs become closer, Alice can establish a stronger communication link with Bob without risking detection by Willie, all while maintaining the quality of public communication between Alice and Carol. However, as the distance between UAVs increases, it becomes increasingly challenging for Alice to transmit data effectively to Bob while simultaneously meeting covert communication constraints. Consequently, optimizing UAV trajectories is highly practical and beneficial in such scenarios.

\section{Conclusion}\label{sec:con}
This paper proposed an F-RIS-aided covert communication framework within UAV networks. We established an electromagnetic model for F-RIS, where the interaction of a parallel beam with the surface is treated as a combination of beams incident at varying angles on individual RIS units. Furthermore, we developed a fitted model that describes the relationship between reflection amplitude, phase, and incident angle, offering valuable insights into the angle-dependent reflection properties of F-RIS. Moreover, to enhance the transmission performance in the proposed network, we introduced NOMA technology for efficient spectrum sharing while maintaining covert constraints. We formulated a complicated non-convex optimization problem, which needs to jointly optimize UAV trajectories, F-RIS reflection vectors, F-RIS incident angle, and power allocation. To address this problem, we utilized a DSAC-based algorithm to learn the strategy for covert communications. Simulation results present the importance of UAV trajectory optimization, adaptive F-RIS reflection phase adjustment, and flexible deformation control. Moreover, we compared and analyzed the proposed algorithm with benchmarks to verify the advantages of the proposed algorithm and framework. Our outcomes indicate that F-RIS has huge potential for secure, adaptable, and efficient covert communications in future UAV networks.

\bibliographystyle{IEEEtran}
\bibliography{ref}
\end{document}